\documentstyle[12pt]{article}
\newcommand{\bb}{\begin{eqnarray}}
\newcommand{\ee}{\end{eqnarray}}
\newcommand{\IR}{ I\!\!R}
\newcommand{\al}{\alpha}
\newcommand{\m}{\mu}
\newcommand{\n}{\nu}
\newcommand{\k}{\kappa}
\newcommand{\be}{\beta}
\newcommand{\g}{\gamma}
\newcommand{\de}{\delta}
\newcommand{\e}{\epsilon}
\newcommand{\si}{\sigma}

\newcommand{\la}{\lambda}
\newcommand{\f}{\phi}
\newcommand{\ep}{\epsilon}

\newcommand{\ff}{\frac}
\newcommand{\nn}{\nonumber}
\newcommand{\pl}{\partial}
 
\begin{document}

\title{\vspace{-1cm} \hfill {\small
hep-th/9506205 }
\vspace{-.5cm}\\
\hfill {\small NTUA-95/51}
\vspace{1cm}\\
{\LARGE INFINITE-DIMENSIONAL ALGEBRAS IN DIMENSIONALLY REDUCED STRING
THEORY}}
\author{Alexandros A. Kehagias \thanks{e-mail: kehagias@sci.kun.nl}
\thanks{Address after September 1995: Phys. Dept., Inst. f\"ur Theor.
Phys., Technische Univ., M\"unchen, D-85748 Garching, Germany}\\
 Physics Department \\
National Technical University\\
15780 Zografou Campus Athens, Greece}
\date{}
\maketitle
\begin{abstract}
We examine 4-dimensional string backgrounds
compactified over a two torus.
There exist two alternative effective
Lagrangians containing each two $SL(2)/U(1)$
sigma-models.
Two of these sigma-models are the complex and K\"ahler
structures on the torus. The
effective Lagrangians are invariant under two different $O(2,2)$
groups and by the successive applications of these groups
 the affine $\widehat{O}(2,2)$ Kac-Moody algebra is emerged.
 The latter has also
a non-zero central term  which generates constant
Weyl rescalings of the reduced 2-dimensional background. In addition,
there exists a number of discrete symmetries relating the field
content of the reduced effective Lagrangians.
\end{abstract}

\newpage

It is known that higher-dimensional gravitational theories exhibit
unexpected new symmetries upon reduction \cite{1}.
  Dimensional reduction of the string
background equations \cite{1'} with dilaton and antisymmetric field also
exhibit
new symmetries as for example dualities \cite{2}--\cite{2''}.
 However, the exact string
symmetries will necessarily be subgroups or discrete versions of the
full symmetry group of the string background equations and thus, a
study of the latter would be useful. The empirical
rule is that the rank of the symmetry group increases by one as the
dimension of the space-time is decreased by one after dimensional
reduction \cite{3}. However, the appearance of non-local currents in
two-dimensions in addition to the local ones, turns the symmetry group
infinite dimensional. Let us recall the $O(8,24)$ group of the
heterotic string after reduction to three dimensions \cite{4}
which turns out to be the
affine $\widehat{O}(8,24)$  algebra by further reduction to two
dimensions \cite{5} or the $\widehat{O}(2,2)$ algebra
after the reduction of
4-dimensional backgrounds \cite{6}.
 The latter generalizes the Geroch group of Einstein
gravity \cite{7}--\cite{7''}.
 We will examine here the ``affinization" of the symmetry
group of the string background equations for 4-dimensional space-times
with two commuting Killing vectors and we will show the emergence of a
central term.
Generalization to higher dimensions is straightforward.

The Geroch group is the symmetry
group which acts  on the space of
solutions of the Einstein equations \cite{7}. Its counterpart in string
theory, the ``string Geroch group",
acts, in full analogy,  on
the space of solutions  of the one-loop beta functions equations
\cite{6}.
The Geroch group, as well as its string counterpart, results by
dimensional reducing four-dimensional backgrounds with zero
cosmological constant over two commuting,
orthogonal transitive, Killing vectors
or, in other words, by compactifing $M_4$ to $M_2\times T^2$. In
dimensional reduced Einstein gravity, there exist two
$SL(2,\IR)$ groups (the Ehlers' and the Matzner-Misner  groups
\cite{8}) acting
on the  space of solutions, the interplay of which produce the
infinite dimensional Geroch group. In the string case, we will see
that apart from the  Ehlers and Matzner-Misner groups acting on the pure
gravitational sector, there also exist two other $SL(2,\IR)$ groups,
 one of which
generates the familiar S-duality,
acting on the antisymmetric-dilaton fields sector.

The Geroch group was also studied in the Kaluza-Klein reduction of
supergravity theories \cite{1}. It was B. Julia  who showed that the Lie
algebra of the Geroch group in Einstein gravity is the affine
Kac-Moody algebra $\widehat{sl}(2)$ and he pointed out the existence
of a central term \cite{8}.
 We will show here that in the string case,
after the reduction to $M_2\times T^2$, there exist four $SL(2,\IR)$
groups, the interplay of which produce the infinite dimensional
Geroch group. However, there is also a central
term which rescales the metric of $M_2$ so that the Lie algebra of
the string Geroch group turns out to be the
$\widehat{sl}(2)\!\times\!\widehat{sl}(2)\simeq
\widehat{o}(2,2)$
affine Kac-Moody algebra. The appearance of a non-zero
 central term already at the tree-level is rather
surprising since usually such terms arise as a concequence
 of  quantization \cite{Ob}. Here however, the
central term acts non-trivially even at the ``classical level" by
constant Weyl rescalings of the reduced two-dimensional space $M_2$.

String propagation in a critical background ${\cal M}$, parametrized
with coordinates $(x^M)$ and metric $G_{MN}(x^M)$, is described by a
two-dimensional sigma-model action
\bb
S=\ff{1}{4\pi\al^\prime}\int d^2z \left( G_{MN}+B_{MN}\right) \pl
x^M\bar{\pl}x^N-\ff{1}{8\pi}\int d^2z \f R^{(2)}\, ,
\ee
where $B_{MN},\, \f$ are  the antisymmetric and dilaton fields,
respectively.
The conditions for conformal invariance at the 1-loop
level in the coupling constant
$\al^\prime$ are
\bb
R_{MN}-\ff{1}{4}H_{MK\Lambda}{H_N}^{K\Lambda}-\nabla_M\nabla_N \f&=&0\,
\nonumber \\
\nabla^M (e^\f H_{MNK})&=&0 \, \nonumber \\
-R+\ff{1}{12}H_{MNK}H^{MNK}+2\nabla^2\f+(\pl_M\f)^2&=&0 \, ,\label{b}
\ee
 and the above equations  may be
 derived from the  Lagrangian \cite{Pe}
\bb
{\cal L}=
\sqrt{-G}
e^\f(R-\ff{1}{12}H_{MNK}H^{MNK}+\pl_M\f\pl^M\f), \label{4}
\ee
where $H_{MN\Lambda}=\pl_M B_{N\Lambda} +cycl.\, perm.$ is the field
strength of the antisymmetric tensor field $B_{MN}$.

 The right-hand
side of the last equation in  eq. (\ref{b})
has been set to zero assuming that the central
charge deficit $\de c$ is of order ${\al^\prime}^2$ (no cosmological
 constant).
 We will also assume that the
string propagates  in $M_4\times K$ with $c(M_4)= 4+{\cal
O}({\al^\prime}^2)$ and that  the dynamics is completely determined by
$M_4$ while the dynamics
of the internal space $K$ is irrelevant for our purposes.
Thus, we will discuss below general 4-dimensional curved
 backgrounds in which
 $H_{\m\n\rho}$ can always be  expressed as the dual of $H^M$
\bb
H_{MN\Lambda}=\ff{1}{2} \sqrt{-G}\, \eta_{MN\Lambda K}H^K \label{5} ,
\ee
with $\eta_{1234}=+1$ and $M,N,...=0,1,2,3$. The
 Bianchi identity $\pl_{[K}H_{MN\Lambda]}=0$ gives the constraint
\bb
\nabla_M H^M=0 \, , \label{6}
\ee
which can be incorporated into (\ref{4}) as $b\nabla_M H^M$
by employing the Lagrange multiplier $b$
so that (\ref{4}) turns out to be
\bb
{\cal L}=
\sqrt{-G}
e^\f(R-\ff{1}{2}sH_M H^M +\e^{-\f}b\nabla_M H^M +
\pl_M\f\pl^M\f). \label{7}
\ee
 $s=\pm 1$ for spaces of Euclidean or Lorentzian signature,
respectively and we will assume that $s=-1$  since the results
may easily be generalized to include the $s=+1$ case as well.
We may now eliminate $H_M$ by using its  equation of motion
\bb
H_M=e^{-\f}\pl_M b\, ,\label{H}
\ee
and  the Lagrangian (\ref{7}) turns out to be
\bb
{\cal L}=
\sqrt{-G}
e^\f(R-\ff{1}{2}e^{-2\f}\pl_M b\pl^M b+
\pl_M\f\pl^M\f). \label{8}
\ee

Let us now suppose that the space-time $M_4$ has an abelian space-like
isometry
generated by the Killing vector ${\bf \xi}_1=\ff{\pl}{\pl\theta_1}$ so that the
metric can be written as
\bb
ds^2=G_{11}d\theta_1^2+2G_{1\m}d\theta_1 dx^\m+G_{\m\n}dx^\m dx^\n
\, , \label{ds}
\ee
where  $\m,\n,...=0,2,3$ and $G_{11},\, G_{1\m},\, G_{\m\n}$ are
functions of $x^\m$.
We may express  the metric (\ref{ds}) as
\bb
ds^2=G_{11}(d\theta_1+2A_\m  dx^\m)^2+\g_{\m\n}dx^\m dx^\n
\, , \label{ds1}
\ee
where \bb
\g_{\m\n}&=&G_{\m\n}-\ff{G_{1\m}G_{1\n}}{G_{11}} \, , \nonumber \\
A_\m&=&\ff{G_{1\m}}{G_{11}}\, .
\ee
 The metric (\ref{ds1}) indicates the
$M_3\times S^1$ topology of $M_4$ and $\g_{\m\n}$ may be considered  as
 the metric of the
3-dimensional space $M_3$. Space-times of this form  have extensively
 been studied in the Kaluza-Klein
reduction where $A_\m$ is considered as a U(1)--gauge field. The scalar
curvature $R$ for the metric (\ref{ds1}) turns out to be
\bb
R=R(\g)-\ff{1}{4}G_{11}F_{\m\n}F^{\m\n}-
\ff{2}{{G_{11}}^{1/2}}\nabla^2{G_{11}}^{1/2}\, ,        \label{Ro}
\ee
where $F_{\m\n}=\pl_\m A_\n-\pl_\n A_\m$ and
$\nabla^2=\ff{1}{\sqrt{-\g}}\pl_\m \sqrt{-\g}\g^{\m\n}\pl_\n$.
By replacing (\ref{Ro}) into (\ref{4}) and integrating by parts we get the
reduced Lagrangian
\bb
{\cal L}=\sqrt{-\g} {G_{11}}^{1/2}e^{\f}\left( R(\g)-\ff{1}{4}G_{11}F_{\m\n}
F^{\m\n}+\ff{1}{G_{11}}\pl_\m
G_{11}\pl^\m\f-\ff{1}{4}\ff{1}{G_{11}}H_{\m\n}H^{\m\n}+\pl_\m\f\pl^\m\f
\right)\label{L}
\ee
where $H_{\m\n}=H_{\m\n1}=\pl_\m B_{\n1}-\pl_\n B_{\m1}$. (A general
discussion on the dimensional reduction of various tensor fields can
be found in \cite{FM}).
 We have taken $H_{\m\n\rho}=0$ since in three dimensions
$B_{\m\n}$ has no physical degrees of freedom.
Let us note that the Lagrangian (\ref{L})  is invariant under
the transformation
\bb
G_{11}&\rightarrow & \ff{1}{G_{11}} \, , \nn \\
H_{\m\n} &\rightarrow & F_{\m\n} \, , \nn \\
\f &\rightarrow & \f-\ln G_{11} \, , \nn \\
\g_{\m\n} &\rightarrow & \g_{\m\n} \, ,
\ee
which, in terms of $G_{11},\,G_{1\m},\,G_{\m\n},\,B_{1\m}$ and $\f$
may be written as
\bb
G_{11}\rightarrow \ff{1}{G_{11}} &,& B_{\m 1}\rightarrow
\ff{G_{\m1}}{G_{11}} \, , \nn \\
G_{1\m}\rightarrow \ff{B_{\m1}}{G_{11}} & ,&
G_{\m\n}\rightarrow G_{\m\n}-\ff{{G_{1\m}}^2-{B_{\m1}}^2}{G_{11}} \,
,\nn \\
\f&\rightarrow& \f-\ln G_{11}  \, ,
\ee
and it  is easily be recognized as
the  abelian duality transformation.

Let us further assume that $M_3$ has also an abelian spece-like  isometry
generated by ${\bf \xi}_2 =\ff{\pl}{\pl\theta_2}$ so that $M_3=M_2\times S^1$.
We
will further assume that the two Killings $({\bf \xi}_1,{\bf \xi}_2)$
 of $M_4$ are
orthogonal to the surface $M_2$. Thus, the metric (\ref{ds})
can be
written as
\bb
ds^2=G_{11}d\theta_1^2
+2G_{12}d\theta_1d\theta_2+G_{22}d\theta_2^2+G_{ij}dx^idx^j \,
,\ee
where $i,j,...=0,3$ and
 $G_{11},\,G_{12},\,G_{22},\, G_{ij}$ are functions of $x^i$
only. We may write the metric above as
\bb
ds^2=G_{11}(d\theta_1+Ad\theta_2)^2+Vd\theta_2^2+G_{ij}dx^idx^j \, ,
\label{sd2}
\ee
where
\bb
A=\ff{G_{12}}{G_{11}} & ,&  V=\ff{G_{11}G_{22}-G_{12}^2}{G_{11}}\,.
\ee
By further reducing (\ref{L}) with respect to ${\bf \xi}_2$ and using the
fact that the only non-vanishing components of $F_{\m\n}$ and
$H_{\m\n}$ are
\bb
F_{i2}&=&\pl_i A \, , \nn \\
H_{i2}&=& \pl_i B \, ,
\ee with $B=B_{21}$, we get
\bb
{\cal L}&=&\sqrt{-G^{(2)}}{G_{11}}^{1/2}V^{1/2}e^{\f}
\left(
R(G^{(2)}) -\ff{1}{2}
(\pl A)^2\ff{G_{11}}{V}-
\ff{1}{8}(\pl \ln \ff{G_{11}}{V})^2
\right. \nn \\
&&\left.-\ff{1}{2}
(\pl B)^2\ff{1}{G_{11}V} -\ff{1}{8}(\pl \ln G_{11}V)^2+ (\pl
\tilde{\f})^2\right) \, \, , \label{L3}
\ee
where $\tilde{\f}=\f+\ff{1}{2}\ln G_{11}V$ and
$(\pl\f)^2=\pl_i\f\pl^i\f$.
Let us now introduce the two complex coordinates $\tau\, , \rho$
\cite{DV} defined by
\bb
\tau=\tau_1+i\tau_2&=&\ff{G_{12}}{G_{11}}+i\ff{\sqrt{G}}{G_{11}}\,,\label{cs}\\
\rho=\rho_1+i\rho_2&=&B_{21}+i\sqrt{G} \, , \label{ks}
\ee
where $G=G_{11}G_{22}-{G_{12}}^2$ is the determinant of the metric on
the 2-torus $T^2=S^1\times S^1$, so that $\tau \, , \rho$ turn out to be
the complex and K\"ahler structure on $T^2$. In terms of $\tau\, ,\rho$,
the Lagrangian (\ref{L3}) is written as
\bb
{\cal L}=\sqrt{-G^{(2)}}e^{\tilde{\f}}&\left( \right.&\left.R(G^{(2)}) +2
\ff{\pl\tau\pl\bar{\tau}}{(\tau-\bar{\tau})^2} +2
\ff{\pl\rho\pl\bar{\rho}}
{(\rho-\bar{\rho})^2}+(\pl\tilde{\f})^2\right)
\, , \label{L4}
\ee
where $R(G^{(2)})$ is the curvature scalar of $M_2$.
The Lagrangian above is clearly invariant under the $SL(2,\IR)\times
SL(2,\IR)\simeq O(2,2,\IR)$ transformation
\bb
\tau&\rightarrow& \tau^\prime=\ff{a\tau+b}{c\tau+d}\,\,\,\,
 , \,\,\,ad-bc=1, \nonumber \\
\rho&\rightarrow&\rho^\prime=\ff{\al\rho+\be}{\g\rho+\de}\,\,\, , \,\,\,
 \al\de-\g\be=1 \,. \label{SS}
\ee
There also exist discrete symmetries acting on the $(\tau,\,\rho)$ space
which leave $\tilde\f$ invariant. One of these interchanges the
complex and K\"ahler structures
\bb
D:&\tau\leftrightarrow\rho& \, ,\tilde{\f}\rightarrow\tilde{\f}
\, . \label{D} \ee
In terms of the fields $G_{11}\, , G_{12}\, ,G_{22}$, and $B_{12}$ the
above transformation is written as
\bb
G_{11}\stackrel{D}{\rightarrow} \ff{1}{G_{11}} &,& G_{12}
\stackrel{D}{\rightarrow}
\ff{B_{21}}{G_{11}} \, , \nn \\
B_{21}\stackrel{D}{\rightarrow} \ff{G_{12}}{G_{11}} & ,&
G_{22}\stackrel{D}{\rightarrow} G_{22}-\ff{{G_{12}}^2-{B_{21}}^2}{G_{11}}
\, ,
\ee
which may be recognized as the factorized duality.

Other discrete symmetries are \cite{2'}
\bb
W:&(\tau,\rho)\leftrightarrow& (\tau,-\bar{\rho})\,\,\, ,
\tilde{\f}\rightarrow\tilde{\f}
\, , \label{W}
\ee
as well as
\bb
R:&(\tau,\rho)\leftrightarrow &(-\bar{\tau},\rho)\,\,\, ,
\tilde{\f}\rightarrow\tilde{\f}
\, , \label{R}
\ee
with $R=DWDW$.
The $W\, , R$ discrete symmetries leave invariant the fields
$G_{ij},\, G_{11},\, G_{22}$ and $\f$ while
\bb
G_{12}\stackrel{W}{\rightarrow}
G_{12}&,&B_{21}\stackrel{W}{\rightarrow} -B_{21} \, , \nonumber \\
G_{12}\stackrel{R}{\rightarrow}
-G_{12}&,&B_{21}\stackrel{R}{\rightarrow} -B_{21} \, .
\ee

Let us note that there exists
another Lagrangian which leads to the same equations as (\ref{L4}). In
can be constructed by using the fact that in 3-dimensions, two-forms
like $F_{\m\n}$ and $H_{\m\n}$ can be  written as
\bb
F^{\m\n}&=&\ff{1}{\sqrt{3}}\ff{\eta^{\m\n\rho}}{\sqrt{-\g}}F_\rho \,
,\nonumber  \\
H^{\m\n}&=&\ff{1}{\sqrt{3}}\ff{\eta^{\m\n\rho}}{\sqrt{-\g}}H_\rho \, .
\ee
The Bianchi identities for $F_{\m\n},\, H_{\m\n}$ are then imply
\bb
\nabla_\m F^\m=0&,&\nabla_\m H^\m=0 . \label{con}
\ee
Thus, we may express (\ref{L}) as
\bb
{\cal L^{*}}&=&
\sqrt{-\g}
{G_{11}}^{1/2}e^\f\left( R+
\ff{1}{2}G_{11}F_\m F^\m +{G_{11}}^{-1/2}\e^{-\f}\psi \nabla_\m F^\m
\right.\\ \nonumber  &&\left. +
\ff{1}{2}\ff{1}{G_{11}}H_\m H_\m +{G_{11}}^{-1/2}\e^{-\f}b\nabla_\m H^\m +
\pl_\m\f\pl^\m\f\right), \label{7'}
\ee
where the constraints
(\ref{con}) have been taken into account by employing the auxiliary
fields $(b,\, \psi)$. The equations of motions for the $H_\m,\, F_\m$
give
\bb
F_\m&=&{G_{11}}^{-3/2}e^{-\f}\pl_\m \psi \, , \nonumber \\
H_\m&=&{G_{11}}^{1/2}e^{-\f}\pl_\m b \, ,
\ee
so that  ${\cal L^{*}}$ is written as
\bb
{\cal
L^{*}}&=&\sqrt{-\g}{G_{11}}^{1/2}e^{\f}\left( R(\g)
-\ff{1}{2}\ff{1}{{G_{11}}^2}e^{-2\f}\pl_\m
\psi \pl^\m \psi \right.
\nonumber  \\
 &&\left.-\ff{1}{2}e^{-2\f}\pl_\m b\pl^\m b+\pl_\m \f \pl^\m
\f\right)
\, . \label{4'}
\ee
If we further reduce it
with respect to ${\bf \xi}_2$, we get
\bb
{\cal
L^{*}}=\sqrt{-G^{(2)}}{G_{11}}^{1/2}V^{1/2}
e^{\f}&\!\left(\right.\!&R(G^{(2)})+\ff{1}{2}\ff{\pl V}{V}\ff{\pl
G_{11}}{G_{11}}-\ff{1}{2}\ff{1}{{G_{11}}^{2}}
e^{-2\f}(\pl \psi)^2
  \nonumber\\ &&\left.+\ff{1}{2}e^{-2\f}(\pl b)^2+(\pl \f)^2
\right)
\, . \label{E1}
\ee
The  two Lagrangians
 $\cal L$, ${\cal L^{*}}$  given by (\ref{L3}) (or
(\ref{L4}))
and  (\ref{E1}), respectively lead to the same equations of motions.
 ${\cal L}$
is invariant under $SL(2,\IR)\!\times\!SL(2,\IR)$ while the symmetries
of $\cal L^{*}$ are less obvious.
In order the invariance properties of both $\cal L$, $\cal
L^{*}$ to become transparent, we adapt the following
parametrization
\bb
G_{11}=e^{-\f}\si\, &,&V=e^{-\f}\ff{\m^2}{\si} \, \\
G_{ij}=e^{-\f}\ff{\la^2}{\si}\eta_{ij}\, &,&
\ee
where $\eta_{ij}=(-1,1)$. The metric
(\ref{sd2}) is then written as
\bb
ds^2=e^{-\f}\si(d\theta_1+Ad\theta_2)^2+e^{-\f}\ff{1}{\si}(\m^2d\theta_2^2
+\la^2\eta_{ij}dx^idx^j)\, .
\ee
As a result,
 $\cal L$, $\cal L^{*}$ turn out to be
\bb
{\cal L}&=&\m\left(
 2\pl \ln\m \pl(\ff{e^{-\f/2}\la\m^{1/2}}{\si^{1/2}})
-\ff{1}{2}\ff{\si^2}{\m^2}(\pl A)^2-\ff{1}{2}(\pl
\ln\ff{\si}{\m})^2\right. \\  \nonumber
&&\left. -\ff{1}{2}\ff{e^{2\f}}{\m^2}(\pl B)^2-\ff{1}{2}(\pl \ln
e^{-\f}\m)^2\right)\, , \label{MM}
\ee
and
\bb
{\cal L^{*}}=\m \left(2\pl\m
\pl\ln\la
-\ff{1}{2}\ff{1}{\si^2}(\pl\si)^2-\ff{1}{2}\ff{1}{\si^2}(\pl\psi)^2
-\ff{1}{2}(\pl\f)^2-\ff{1}{2}e^{-2\f}(\pl b)^2\right)\, . \label{Eh}
\ee
Note that $(A,\psi)$ and $(B,b)$
are related through the relations
\bb
\pl_iA&=&-\ff{1}{\sqrt{3}}\varepsilon_{ij}\ff{\m}{\si^2}\eta^{jk}\pl_k\psi
\, ,\label{AP}\\
\pl_iB&=& -\ff{1}{\sqrt{3}}\varepsilon_{ij}e^{-2\f}\m\eta^{jk}\pl_kb\,
, \label{Bb}\ee
where $\varepsilon_{12}=1$ is the antisymmetric symbol in
two-dimensions.

Let us now define, in addition to the $(\tau,\, \rho)$ fields
given in eqs. (\ref{cs},\ref{ks}), the  complex fields $(S,\, \Sigma)$
\bb
S=b+ie^{\f}&,& \Sigma=\psi+i\si\, .
\ee
Then ${\cal L, L^{*}}$ may be expressed as
\bb
{\cal L}&=&\m\left(
 2\pl \ln\m \pl(\ff{e^{-\f/2}\la\m^{1/2}}{\si^{1/2}})
 +2\ff{\pl \tau\pl\bar{\tau}}{(\tau-\bar{\tau})^2}+2
\ff{\pl\rho\pl\bar{\rho}}
{(\rho-\bar{\rho})^2}\right) \\
{\cal L^{*}}&=&\m \left(2\pl\m
\pl\ln\la
 +2\ff{\pl S\pl\bar{S}}{(S-\bar{S})^2}+2 \ff{\pl\Sigma\pl\bar{\Sigma}}
{(\Sigma-\bar{\Sigma})^2}\right) \, .
\ee
Thus, there exist four  $SL(2,\IR)/U(1)$--sigma models,
${\cal L}$ is invariant under the $SL(2,\IR)\!\times\!SL(2,\IR)$
transformations (\ref{SS})
and  ${\cal L^{*}}$ is invariant under
\bb
S\rightarrow \ff{kS+m}{nS+\ell} &,& \Sigma\rightarrow
\ff{\k\Sigma+\eta}{\n\Sigma+\theta}\, .
\ee
These transformation do not affect $\m$. There also exist discrete
$Z_2$ transformations, besides those that have already been noticed in
eqs. (\ref{D},\ref{W},\ref{R}), namely
\bb
D^\prime: (S,\Sigma) &\leftrightarrow & (\Sigma,S) \\
W^\prime: (S,\Sigma) &\leftrightarrow & (S,-\bar{\Sigma}) \\
R^\prime: (S,\Sigma) &\leftrightarrow & (-\bar{S},\Sigma)\, .
\ee
Moreover, the transformations
\bb
N: (\tau,\rho)\leftrightarrow (S,\Sigma) &,& \la \leftrightarrow
e^{-\f/2}
\ff{\m^{1/2}}{\si^{1/2}}\la \, ,
\\
N^\prime: (\tau,\rho)\leftrightarrow (\Sigma,S) &,& \la \leftrightarrow
e^{-\f/2}
\ff{\m^{1/2}}{\si^{1/2}}\la \, ,
\ee
indentify the two Lagrangians and thus, may be considered as the
string counterpart of the Kramer-Neugebauer symmetry \cite{KN}.
Note  that  $\cal L, L^{*}$ may also be written as
\bb
{\cal L}&=& \m \left(
 2\pl \ln\m \pl(\ff{e^{-\f/2}\la\m^{1/2}}{\si^{1/2}})
-\ff{1}{4} Tr({h_1}^{-1}\pl h_1)^2-\ff{1}{4} Tr({h_2}^{-1}\pl h_2)^2\right) \\
{\cal L^{*}}&=&\m \left( 2\pl \m \pl\la -\ff{1}{4} Tr({g_1}^{-1}\pl
g_1)^2-\ff{1}{4} Tr({g_2}^{-1}\pl g_2)^2\right) .
\ee
where the $2\!\times\!2$ matrices $h_1,\, h_2,\, g_1$ and $g_2$ are
\bb
h_1=\left( \begin{array}{cc}
\ff{\si}{\m}& \ff{\si}{\m}A \\
\ff{\si}{\m}A&\ff{\si}{\m}A^2+\ff{\m}{\si}\end{array}\right) &,&
h_2=\left( \begin{array}{cc}
\ff{e^{\f}}{\m}& \ff{e^{\f}}{\m}B \\
\ff{e^{\f}}{\m}B&\ff{e^{\f}}{\m}B^2+\ff{\m}{e^{\f}}\end{array}\right)
 \, ,\\
g_1=\left( \begin{array}{cc}
\ff{1}{\si}& \ff{1}{\si}\psi \\
\ff{1}{\si}\psi&\ff{1}{\si}\psi^2+\si\end{array} \right) &,&
g_2=\left( \begin{array}{cc}
e^{-\f}& e^{-\f}b \\
e^{\f}b&e^{\f}b^2+e^{-\f}\end{array}\right) \, .
\ee
The Lagrangian ${\cal L}$ is invariant under the infinitesimal
transformations
\bb
\de \si=\sqrt{2}\ff{1}{\si}A\ep_1^+-2\ep_1^0 &,&
\de A=
-\frac{1}{\sqrt{2}}(\ff{\si^2}{\m^2}-A^2)\ep_1^+-2A\ep_1^0+\sqrt{2}\ep_1^-\, ,
\nonumber \\
\de \f =-\sqrt{2} B\ep_2^++2 \ep_2^0 &,&
\de B=
-\frac{1}{\sqrt{2}}(\ff{e^{2\f}}{\m^2}-B^2)\ep_2^+-2B\ep_2^0+\sqrt{2}\ep_2^-
\, ,  \label{tr1}
\ee
while ${\cal L^{*}}$ is invariant under
\bb
\de \si =-\sqrt{2}\psi\si \ep_3^++2\si\ep_3^0 &,&
\de \psi=
-\frac{1}{\sqrt{2}}(\ff{1}{\si^2}-\psi^2)\ep_3^+
-2\psi\ep_3^0+\sqrt{2}\ep_3^-\,
,
\nonumber \\
\de \f  =\sqrt{2}b\ep_4^+-2\ep_4^0 &,&
\de b=
-\frac{1}{\sqrt{2}}(e^{2\f}-b^2)\ep_4^+-2b\ep_4^0+\sqrt{2}\ep_4^-
\, .  \label{tr2}
\ee
The above infinitesimal transformations are generated by a set of
 four Killing
vectors $({\bf K}_a^{(i)},\,a=1,2,3,\, i=1,2,3,4)$ which can easily be
written down by recalling that the
metric
\bb
ds^2=dx^2+e^{2x}dy^2
\ee
has a three-parameter group of isometries generated by
\bb
K_+&=&-\sqrt{2} y\pl_x -\ff{1}{\sqrt{2}}(e^{-2x}-y^2)\pl_y \, ,
\nonumber \\
K_0&=&2(\pl_x-y\pl_y) \, , \nonumber \\
K_-&=&\sqrt{2}\pl_y \, ,
\ee
which satisfy the $SL(2)$ commutation relations
\bb
 [K_+,K_0]=2K_+\, \, \, ,[K_-,K_0]=-2K_-\, \, \, , [K_-,K_+]=-K_0\, .
\ee
Among these Killing vectors, let us consider $K_0^{(3)}$ which
scales both  $\psi$ and $\si$ as
\bb
K_0^{(3)}:\, \,
(\psi,\si)\rightarrow (\al\psi,\al\si).
\ee
In view of eq. (\ref{AP}), $A$ is also scaled as
\bb
A\rightarrow \ff{1}{\al} A\, ,
\ee
so that $(A,\si)$ is transformed into
$(\ff{1}{\al}A,\al\si)$ which is
generated by $-K_0^{(1)}$. However, ${\cal L}$ is not
invariant unless we also scale the conformal factor $\la$ as
$\sqrt{\al}\la$. Let us denote the generator of constant Weyl
transformations by $k$. Then we have the relation
\bb
K_0^{(1)}+K_0^{(3)}=k \, .
\ee
In the same way, one may see that $K_0^{(2)},\,K_0^{(4)}$ which
transform $(B,\f)$ and $(b,\f)$ as $(e^{-\al} B,\f+\al)$,
$(e^\al,\f+\al)$
respectively satisfy
\bb
K_0^{(2)}+K_0^{(4)}=k.
\ee
As a result, the algebra turns out to be
\bb
\begin{array}{lll}
 [K_+^{(1)},K_0^{(1)}]=2K_+^{(1)}\, ,&[K_-^{(1)},K_0^{(1)}]=
-2K_-^{(1)}\,, & [K_-^{(1)},K_+^{(1)}]=K_0^{(1)}\, , \\
 {}[K_+^{(2)},K_0^{(2)}]=2K_+^{(2)} \, , &
[K_-^{(2)},K_0^{(2)}]=-2K_-^{(2)}\,  , &
[K_-^{(2)},K_+^{(2)}]=K_0^{(2)}\, ,  \\
{}[K_+^{(3)},k\!-\!K_0^{(1)}]=2K_+^{(3)} \, ,&
[K_-^{(3)},k\!-\!K_0^{(1)}]=-2K_-^{(3)}\, , &
[K_-^{(3)},K_+^{(3)}]=k\!-\!K_0^{(1)}\, , \\
{} [K_+^{(4)},k\!-\!K_0^{(2)}]=2K_+^{(4)} \, ,
& [K_-^{(4)},k\!-\!K_0^{(2)}]=-2K_-^{(4)} \, , &
[K_-^{(4)},K_+^{(4)}]=k\!-\!K_0^{(2)}\, \end{array} \label{km}
\ee
If we define the generators $(h_i,\,k_i,\,f_i)$ by
\bb
h_i=K_0^{(i)}\, ,\,\, f_i=K_+^{(i)}\, ,\,\, e_i=K_-^{(i)}\, ,
\ee
then the algebra
(\ref{km}) may be writen as
\bb
[h_i,h_j]&=&0\, , \nonumber \\
{}[h_i,e_j]&=&A_{ij}e_j \, , \nonumber \\
{}[h_i,f_j]&=&-A_{ij} \, ,\nonumber \\
{}[e_i,f_j]&=& \de_{ij}h_j \, ,
\ee
where the Cartan martix $A_{ij}$ is
\bb
A_{ij}=\left(\begin{array}{cc}
a_{ij}&0\\
0&a_{ij}
\end{array}\right) &,& a_{ij}=\left(\begin{array}{cc}
2&-2\\
-2&2 \end{array}
\right)\, .
\ee
In addition, one may verify the Serre relation
\bb
(ad e_i)^{1-A_{ij}}(e_j)=0&,&
(ad f_i)^{1-A_{ij}}(f_j)=0\, .
\ee
As a result, the algebra generated by the successive applications of
the transformations (\ref{tr1},\ref{tr2}) is the affine Kac-Moody
algebra $\widehat{o}(2,2)$ with a central term corresponding to
constant Weyl rescalings of the 2-dimensional background metric. The
central term survives in
higher dimensions as well, since its emergence
is related to the existence of two alternative
effective Lagrangians  after reducing the 3-dimensional theory
down to two dimensions over an abelian
isometry.
It is the interplay of the symmetries of these Lagrangians which
produce the  Kac-Moody algebra.

\end{document}